\documentclass[default]{svmult}

\usepackage{mathptmx}       
\usepackage{helvet}         
\usepackage{courier}        
\usepackage{type1cm}        
\usepackage{amsmath}                            
\usepackage{amsbsy}
\usepackage[numbers]{natbib}

\usepackage{makeidx}         
\usepackage{graphicx}        
\usepackage{multicol}        
\usepackage[]{footmisc}

\RequirePackage[OT1]{fontenc}
\RequirePackage{graphicx,subfigure,latexsym,amssymb, url}
\RequirePackage{float,epsfig,multirow,rotating}
\usepackage{algorithm}
\usepackage[noend]{algpseudocode}
\usepackage{xcolor}
\usepackage{gensymb} 
\usepackage[]{inputenc} 
\usepackage{textcomp}
\usepackage{listings}
\usepackage{color}
\RequirePackage[colorlinks,citecolor=blue,urlcolor=blue]{hyperref}
\usepackage{bm}

\numberwithin{equation}{section}

\newcommand{\bof}{\boldsymbol{f}}

\newcommand{\brho}{\boldsymbol{\rho}}

\newcommand{\bxi}{\boldsymbol{\xi}}

\newcommand{\bD}{\boldsymbol{D}}

\newcommand{\bV}{\boldsymbol{V}}

\newcommand{\bS}{\boldsymbol{S}}

\newcommand{\bx}{\boldsymbol{x}}
\newcommand{\bX}{\boldsymbol{X}}
\newcommand{\by}{\boldsymbol{y}}
\newcommand{\bY}{\boldsymbol{Y}}

\makeindex             

\begin{document}

\title*{Learning in the Absence of Training Data -- a Galactic Application}
\titlerunning{Learning without Training Data}
\author{Cedric Spire, Dalia Chakrabarty}
\institute{Cedric Spire \at Loughborough University. Department of
  Mathematical Sciences. \email{c.spire@lboro.ac.uk}
\and Dalia Chakrabarty \at Loughborough University. Department of Mathematical
Sciences. \email{d.chakrabarty@lboro.ac.uk}}
%
%
\maketitle

\abstract*{ There are multiple real-world problems in which training data is
  unavailable, and still, the ambition is to learn values of the system
  parameters, at which test data on an observable is realised, subsequent to
  the learning of the functional relationship between these variables. We
  present a novel Bayesian method to deal with such a problem, in which we
  learn a system function of a stationary dynamical system, for which
  only test data on a vector-valued observable is available, and training data
  is unavailable. This exercise borrows heavily from the state space
  probability density function ($pdf$), that we also learn. As there is no
  training data available for either sought function, we cannot learn its
  correlation structure, and instead, perform inference (using
  Metropolis-within-Gibbs), on the discretised form of the sought system
  function and of the ${pdf}$, where this \textit{pdf} is constructed such
  that the unknown system parameters are embedded within its
  support. Likelihood of the unknowns given the available data, is defined in
  terms of such a \textit{pdf}. We make an application to the
  learning of the density of all gravitational matter in a real galaxy.  }

\abstract{
There are multiple real-world problems in which training data is
  unavailable, and still, the ambition is to learn values of the system
  parameters, at which test data on an observable is realised, subsequent to
  the learning of the functional relationship between these variables. We
  present a novel Bayesian method to deal with such a problem, in which we
  learn a system function of a stationary dynamical system, for which
  only test data on a vector-valued observable is available, and training data
  is unavailable. This exercise borrows heavily from the state space
  probability density function ($pdf$), that we also learn. As there is no
  training data available for either sought function, we cannot learn its
  correlation structure, and instead, perform inference (using
  Metropolis-within-Gibbs), on the discretised form of the sought system
  function and of the ${pdf}$, where this \textit{pdf} is constructed such
  that the unknown system parameters are embedded within its
  support. Likelihood of the unknowns given the available data, is defined in
  terms of such a \textit{pdf}. We make an application to the
  learning of the density of all gravitational matter in a real galaxy.
}

\section{Introduction}
\label{sec:1}
The study of rich correlation structures of high-dimensional random objects, is often invoked when
learning the unknown functional relationship between an observed
random variable, and some other parameters that might inform on the properties
of a system. A problem in which a vector of system parameters (say
$\brho\in{\cal R}\subseteq{\mathbb R}^p$) is related to an observed response
variable (say $\bY\in{\cal Y}\subseteq{\mathbb R}^d$), is easily visualised by
the equation: $\bY=\bxi(\brho)$, where $\bxi:{\cal
  R}\longrightarrow{\cal Y}$. Given training data $\mathbf{D} =
\{(\brho_i,\by_i)\}_{i=1}^{N_{data}}$, we aim to learn this unknown mapping
$\bxi(\cdot)$ within the paradigm of \textit{supervised learning}.
By ''training data'' we mean here the pairs composed of chosen design points
$\brho_i$, and the output $\by_i$ that is generated at $\brho_i$;
$i=1,\ldots,N_{data}$. Methods to perform supervised learning are extensively
covered in the literature \cite{elementML,neal,rasmussen,russelnorvig}. Having learnt
$\bxi(\cdot)$, one could use this model to predict the value $\brho$
\cite{ejs}, at which the test datum $\by_{test}$ on $\bY$ is realised --
either in the conventional framework as
$\brho=\bxi^{-1}(\bY)\vert_{\bY=\by_{test}}$, or as the Bayesian equivalent. Such prediction is possible,
only subsequent to the learning
of the functional relation between $\brho$ and $\bY$ using training data $\mathbf{D}$.

However, there exist physical systems for which only measurements on the
observable $\bY$ are known, i.e. training data is not available. The
disciplines affected by the absence of training data are diverse. In
engineering \cite{Sun2011}, anomaly detection is entirely
sample-specific. There is no training data that allows for the learning of a
functional relationship between anomaly occurrence (parametrised by type and
severity of anomaly) and conditions that the sample is subjected to. Yet, we
need to predict those anomalies. In finance, such anomalies in stock price
trends are again outside the domain of supervised learning, given that the
relationship between the market conditions and prices have not been reliably captured
by any "models" yet. In neuroscience, \cite{ahmad2017}, a series of neurons spike at
different amplitudes, and for different time widths, to cause a response (to a
stimulus). We can measure the response's strength and the parameters of firing
neurons, but do not know the relation between these variables.
Again, in petrophysics, the system property that is the proportion of the
different components of a rock (eg. water, hydrocarbons), affects Nuclear
Magnetic Resonance (NMR) measurements from the rock \cite{6,7}. However, this
compositional signature cannot be reliably estimated given such data, using
available estimation techniques. Quantification of petrological composition
using the destructive testing of a rock, is highly exclusive, and expensive,
to allow for a sample that is large enough to form a meaningful training data
set. Also, the resulting training data will in general be unrepresentative of
any new rock, since the relationship between the (compositional) system
property and (NMR) observable is highly rock-specific, being driven by
geological influences on the well that the given rock is obtained
from. Therefore any training data will need to be substantially diverse, and
as stated before, this is unachievable in general. Equally, this dependence
on the latent geological influence annuls the possibility of using numerical
simulations to generate NMR data, given design compositional
information. Thus, generation of training data is disallowed in general.

In this work, we advance the learning of the sought functional relation
between an observable and a system parameter vector, in such a challenging
(absent training) data situation; this could in principle, then be undertaken
as an exercise in {\it unsupervised learning}, though opting for the more
robust supervised learning route is still possible, as long as the missing
training data is generated, i.e. we are able to generate the $\brho_i$ at which
the measured (test) datum, $\by_i$ on $\bY$, is available, $\forall
i\in\{1,\ldots,N_{data}\}$. Our new method for accomplishing this, is to invoke
a system property that helps link $\brho$ with $\bY$, and this is
possible in physical systems for which we have -- at least partial
-- observed information. To clarify, what we advance in the face of the absent
training data, is the pursuit of the probability density function of the
observable $\bY$, on which data is available, and employ this to
learn the system parameter vector $\brho$. We undertake such an exercise
in a Bayesian framework, in which we seek posterior of the
\textit{pdf} of the observables, and the system parameters, given the
available data. 

The sought parameter vector could inform on the behaviour, or structure, of
the system (eg. it could be the vectorised version of the density function of all gravitating matter in a distant galaxy). The
state space \textit{pdf} establishes the link between this unknown vector, and
measurements available on the observable (that may comprise complete or
incomplete information on the state space variable). We consider
dynamical systems, s.t. the system at hand is governed by a kinetic equation
\cite{kinetic}; we treat the unknown system parameter vector as the stationary
parameter in the model of this dynamical system. In the novel Bayesian
learning method that we introduce, this parameter is embedded within the
support of the state space \textit{pdf}. We describe the general model in
Section~\ref{sec:2}, that is subsequently applied to an astronomical
application that is discussed in Section~\ref{sec:3}. Inference is discussed
in Section~\ref{sec:4}, where inference is made on the state space
{\it pdf} and the sought system parameters, given the data that comprises
measurements of the observable, using Metropolis-within-Gibbs. Results are
presented in Section~\ref{sec:5}, and the paper is rounded up with a
conclusive section (Section~\ref{sec:6}).

\section{General Methodology}
\label{sec:2}
\noindent
We model the system as a dynamical one, and define the state space variable as
a $p$-dimensional vector $\bS\in{\cal S}\subseteq{\mathbb R}^p$. Let the
observable be $\bY\in{\cal Y}\subseteq{\mathbb R}^d;\:d<p$, such that only
some ($d$) of the $p$ different components of the state space vector $\bS$ can
be observed. In light of this situation that is marked by incomplete
information, we need to review our earlier declaration of interest in the
probability density function of the full state space vector. Indeed, we aim to
learn the \textit{pdf} of the state space variable $\bS$, and yet, have
measured information on only $\bY$, i.e. on only $d$ of the $p$ components of
$\bS$. Our data is then one set of measurements of the observable 
$\bY$, and can be expressed by $\bD=\{\by^{(k)}\}_{k=1}^{N_{data}}$. If the
density of ${\cal S}$ is to be learnt given data on $\bY$, such incompleteness
in measured information will have to be compensated for by invoking some
independent information. Such independent information is on the symmetry of ${\cal S}$. 

It follows that unobserved components of $\bS$ will have to be integrated out of the state space \textit{pdf}, in order to compare against data that comprises measurements of the observables. This state space {\it pdf} that the unobserved variables are integrated out of, is equivalently projected onto the space ${\cal Y}$ of observables, and therefore, we refer to it as the \textit{projected state space pdf}. The likelihood of the model parameters, given the data, is simply the product of the projected state space \textit{pdf} over all the data points. But until now, the unknown model parameters have not yet appeared in our expression of the likelihood. The next step is then to find a way for embedding the sought system parameters, in the support of the projected state space \textit{pdf}.

This can be achieved by assuming that our dynamical system is
stationary, so that its state space \textit{pdf} does not depend on
time-dependent variables. In other words, the rate of change of the
state space \textit{pdf} is $0$. This allows us to express the
\textit{pdf} as dependent on the state space vector $\bS$, but only
via such functions of (some or all amongst) $S_1,\ldots, S_p$ that are
not changing with time; in fact, the converse of this statement is
also true. This is a standard result, often referred to as Jeans
Theorem \cite{BT, Jeans}. The model parameters that we seek, can be recast
as related to such identified time-independent functions of
all/some state space coordinates of motion. Thus, by expressing the state
space \textit{pdf} as a function of appropriate constants of
motion, we can embed system parameters into the
support of the sought \textit{pdf}.

As stated above, this \textit{pdf} will then need to be projected into
the space of observables ${\cal Y}$, and we will convolve such a
projected \textit{pdf} with the error density, at every choice of the model parameters. Then assuming the data to be $iid$, the product of
such a convolution over the whole data set will finally define our
likelihood. Using this likelihood, along with
appropriate priors, we then define the posterior probability density of the model parameters and the state space $pdf$, given the data $\bD$. Subsequently we generate 
posterior samples using {Metropolis-within-Gibbs}.
scheme.

We recall that in absence of training data on a pair of r.v.s, we cannot
learn the correlation structure of the functional relationship between these
variables. In such situations, instead of the full function, we can only
learn the vectorised version of the sought function. In other
words, the relevant interval of the domain of the function is
discretised into a bin, and the value of the function held a constant over any
such bin; we can learn the functional value over any such bin.
\section{Astrophysics Application}
\label{sec:3}
\noindent
Our astrophysics application is motivated by the wish to learn the
contribution of dark matter, to the density function of all gravitating mass
in a distant galaxy. While information on light-emitting matter is available,
it is
more challenging to model the effects of dark matter since, by definition, one
cannot observe such matter (as it does not emit/reflect light of any
colour). However, physical phenomena such as: the distortion of the path of
light by gravitational matter acting as gravitational lenses; temperature
distribution of hot gas that is emanating from a galaxy; motions of stars or
other galactic particles that is permitted in spite of the attractive
gravitational pull of the surrounding galactic matter, allow us to confirm
that non-observable, dark matter is contributing to the overall gravitational
mass density of the galaxy. In fact, astrophysical theories suggest that the
proportion of dark matter in older galaxies (that are of interest to us here)
is the major contributor to the galactic mass, over the minor fraction of
luminous galactic matter \cite{veselina_kalinova}. We can compute the
proportion of this contribution, by subtracting the density of
the luminous matter from the overall density. It is then necessary to learn the
gravitational mass density of the whole system in order to learn the density
of dark matter.

We begin by considering the galaxy at hand to be a stationary dynamical
system, i.e. the distribution of the state space variable does not
depend on time. Let $\bS = (X_1,X_2,X_3,V_1,V_2,V_3)^T\in{\cal S}\subseteq{\mathbb{R}^6}$ define the state space variable of a galactic particle, where $\bX=(X_1,X_2,X_3)^T$ is defined as its 3-dimensional location vector and $\bV=(V_1,V_2,V_3)^T$ as the 3-dimensional velocity vector of the galactic particle. Our data consists of measurement of the one observable velocity coordinate $V_3$, and two observable spatial coordinates, $X_1, X_2$, of $N_{data}$ galactic particles (eg. stars). That is, for each galactic particle, we have measurements of $\bY = (X_1,X_2,V_3)^T \in{\cal Y}\subseteq{\mathbb{R}^3}$. For $N_{data}$ observations, our data is thus $\bD=\{\by^{(k)}\}_{k=1}^{N_{data}}$.


The system function that we are interested in learning here, is the density
function $\rho(X_1, X_2, X_3)$ of the gravitational mass of all matter in the
considered galaxy, where we assume that this gravitational mass density
$\rho(\cdot)$ is a function of the spatial coordinates $\bX$ only. This system
function does indeed inform on the structure of the galactic system -- for it
tells us about the distribution of matter in the galaxy; it also dictates the
behaviour of particles inside the galaxy, since the gravitational mass density
is deterministically known as a function of the gravitational potential
$\Phi(X_1, X_2, X_3)$ via the Poisson equation ($\nabla^2 \Phi(X_1, X_2, X_3) =
-4\pi G \rho(X_1, X_2, X_3)$, where $G$ is the known Universal Gravitational
constant, and $\nabla^2$ is the Laplacian operator), which is one of the
fundamental equations of Physics \cite{goldstein}. The potential of a system dictates
system dynamics, along with the state space distribution.

Here, we assume that the state space density of this dynamical system does not
vary with time, i.e.
$\displaystyle{\frac{df\left[X_1(t),X_2(t),X_3(t),V_1(t),V_2(t),V_3(t)\right]}{dt}}
= 0$.  This follows from the consideration that within a typical galaxy,
collisions between galactic particles are extremely rare \cite{BT}. We thus
make the assumption of a collisionless system evolving in time according to
the \textit{Collisionless Boltzmann Equation} (CBE) \cite{BT,CBE}. As
motivated above, this allows us to express the state space \textit{pdf} as
dependent on those functions of $X_1,X_2,X_3, V_1,V_2,V_3$ that remain
invariant with time, along any trajectory in the state space $\mathcal{S}$;
such time-invariant constants of motion include energy, momentum, etc. It is a
standard result that the constant of motion that the state space $pdf$ has to
depend on, is the energy $E(X_1,X_2,X_3,\parallel\bV\parallel)$ of a galactic
particle \cite{binney82,contop63}, where $\parallel\cdot\parallel$ represents
the Euclidean norm of a vector. Here, energy is given partly by kinetic energy
that is proportional to $\parallel\bV\parallel^2$, and partly by potential
energy, which by our assumption, is independent of velocities. Secondly, given
that the state space is $6$-dimensional, the number of constants of motion
$\leq$5, in order to let the galactic particle enjoy at least 1 degree of
freedom, i.e. not be fixed in state space \cite{contop63}.

We ease our analysis by assuming that the state space $pdf$ is a
function of energy only. This can be rendered equivalent to designating the
symmetry of isotropy to the state space ${\cal S}$, where isotropy implies
invariance to rotations in this space, i.e. the state space $pdf$ is assumed
to be such a function of $\bX$ and $\bV$, that all orthogonal transformations
of $\bX$ and $\bV$ preserve the state space $pdf$. The simple way to achieve
the equivalence between a isotropic state space $pdf$ and the lone dependence
on energy $E$ of the $pdf$, is to ensure that the gravitation mass density,
(and therefore the gravitational potential) at all points at a
given Euclidean distance from the galactic centre, be the same, i.e. the
distribution of gravitational mass abides by spherical symmetry
s.t. $\rho(\cdot)$ (and therefore $\Phi(\cdot)$) depends on $X_1,X_2,X_3$ via
the Euclidean norm $\parallel\bX\parallel$ of the location vector $\bX$ of a
particle. Then energy $E$ is given as the sum of the
$\parallel\bV\parallel^2$-dependent kinetic energy, and the
$\parallel\bX\parallel$-dependent potential energy. Spherical mass
distribution is not a bad assumption in the central parts of
``elliptical'' galaxies that are of interest for us, as these have a global triaxial
geometry.

To summarise, state space $pdf$ is written as $f(E)$, and we embed
$\brho(\cdot)$ into the support of this state space $pdf$ $f(E)$, by recalling
that energy $E$ is partly the gravitational potential energy $\Phi(\cdot)$
that is deterministically related to the gravitational mass density
$\rho(\cdot)$ through Poisson equation.

As there is no training data available to learn the correlation structure of
the sought functions $\rho(\bX)$ and $f(E)$, we can only learn values of these
functions at specified points in their domains, i.e. learn their vectorised
forms $\brho$ and $\bof$ respectively, where
$\brho:=(\rho_1,...,\rho_{N_X})^T$, with $\rho_i=\rho(\bx)$ for
$\bx\in[\bx_{i-1}, \bx_i]; i=1,\ldots N_x$. The discretised form of $f(E)$
is similarly defined, after discretising the relevant (non-positive)
$E$-values (to indicate that the considered galactic particles are bound to
the galaxy by gravitational attraction), into $N_E$ number of $E$-bins. 
Then in terms of these vectorised versions of the state space $pdf$ likelihood
of the unknown parameters $\rho_1,\ldots\rho_{N_X},f_1,\ldots,f_{N_E}$, given data on the observable $\bY$ is:
\begin{equation}
\ell\left(\brho,\bof|\{\by^{(k)}\}_{k=1}^{N_{data}} \right)
= \prod_{k=1}^{N_{data}} \nu(\by^{(k)},\brho,\bof),
\label{eq:finalL}
\end{equation} 
where $\nu(.)$ is the projected state space \textit{pdf}.

We also require that $\rho_1\geq 0,\ldots\rho_{N_X}\geq 0,f_1\geq
0,\ldots,f_{N_E} \geq 0$, and that $\rho_i\geq
\rho_{i+1},\:i=1,\ldots,N_X-1$. The latter constraint is motivated by how the
mass in a gravitating system (such as a galaxy) is distributed; given that gravity is an attractive force, the stronger
pull on matter closer to the centre of the galaxy, implies that gravitational
mass density should not increase, as we move away from the centre of the
system. These constraints are imposed via the inference that we employ.
\section{Inference}
\label{sec:4}
\noindent
Inference on the unknown
parameters -- that are the components of $\brho$ and $\bof$ -- is undertaken
using Metropolis-within-Gibbs. In the first block update during any iteration,
the $\rho_1,\ldots,\rho_{N_X}$ parameters are updated, and subsequently, the 
$f_1,\ldots, f_{N_E}$ parameters are updated in the 2nd block, at the updated
$\rho$-parameters, given the data $\bD$ that comprises $N_{data}$ measurements
of the observed state space variables $X_1,X_2,V_3$ that are the components of
the observable vector $\bY$.

Imposition of the monotonicity constraint on the $\rho$ parameters,
s.t. $\rho_i \geq \rho_{i+1}$, $i=1,\ldots N_{X}-1$, renders the inference
interesting. We propose $\rho_i$ from a Truncated Normal proposal density
that is left truncated at $\rho_{i+1}$, $\forall i=1,\ldots,N_X-1$, and
propose $\rho_{N_X}$ from a Truncated Normal that is left truncated at 0. The
mean of the proposal density is the current value of the parameter and the
variance is experimentally chosen, as distinct for each
$i\in\{1,\ldots,N_X\}$. Such a proposal density helps to maintain the
non-increasing nature of the $\rho_i$-parameters, with increasing $i$. At the
same time, non-negativity of these parameters is also maintained. We choose
arbitrary seeds for $\rho_1,\ldots,\rho_{N_X}$, and using these as the means,
a Gaussian prior is imposed on each parameter. The variance of the prior
densities is kept quite large, and demonstration of lack of sensitivity to the
prior choices, as well as the seeds, is undertaken.   
\vspace{-.4cm}
\begin{figure}[!h]
\centering
   \includegraphics[height=6.5cm]{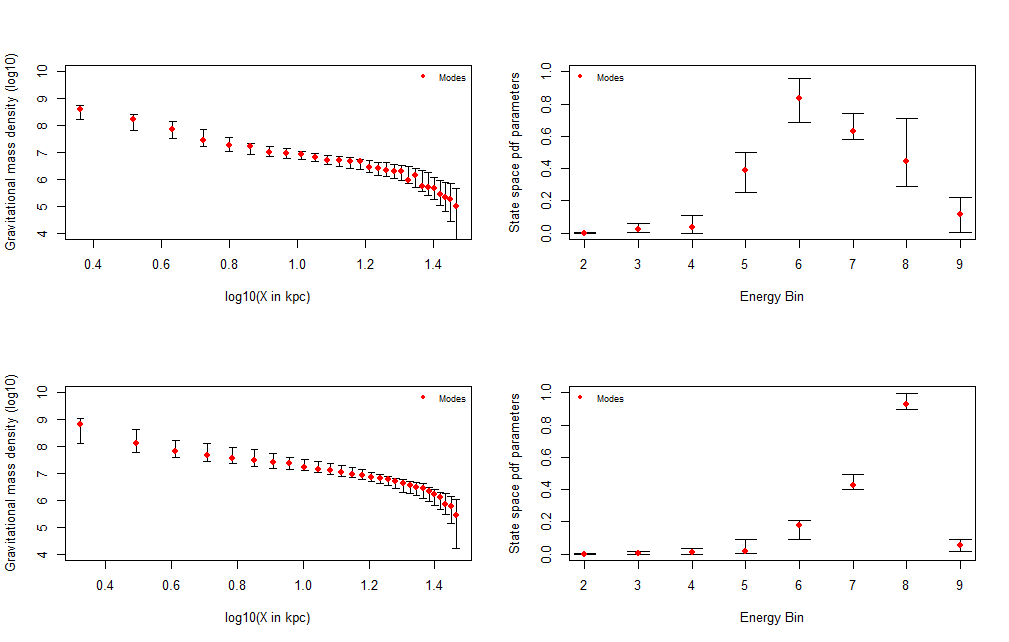} 
\caption{Results from a MCMC scheme showing the $95\%$ HPDs for all the parameters to learn, for both PNe (top row) and GC (bottom row) data. Modes are shown as red dots.\textbf{Top Row:} HPDs on the $\brho$ (left), and the $\bof$ parameters for the PNe data. \textbf{Bottom Row:} HPDs on the $\brho$ (left), and the $\bof$ parameters for the GC data. }
\label{fig:HPDs}
\end{figure} 
\vspace{-.4cm}
As for components of the vectorised state space $pdf$,
there is no correlation information to be enjoyed in this case, unlike in the
case of the components of the vectorised gravitational mass density function.
We propose $f_j$ from a Truncated Normal (to maintain non-negativity), where
the mean of this proposal density is the current value of the parameter and the
variance is chosen by hand. Loose Gaussian priors are imposed, while
the same seed value is used $\forall j\in\{1,\ldots,N_{E}\}$. 
 
An important consideration in our work is the choice of $N_X$ and
$N_{E}$. We could have treated these as unknowns and attempted learning
these from the data; however, that would imply that the number
of unknowns is varying from one iteration to another, and we desired
to avoid such a complication, especially since the data strongly
suggests values of $N_X$ and $N_E$. We choose $N_X$ by binning the range
of $R_p:=\sqrt{X_1^2+X_2^2}$ values in the data $\bD$, s.t. each resulting $R_p$-bin includes at least one
observed value of $V_3$ in it, and at the same time, the number of
$R_p$-bins is maximised. Again, we use the available data $\bD$ to
compute the empirical values of energy $E$, where an arbitrarily scaled
histogram of the observed $R_p$ is used to mimic the vectorised
gravitational mass density function, that is then employed to compute
the empirical estimate of the vectorised gravitational potential
function, that contributes to $E$ values. We admit maximal 
$E$-bins over the range of the empirically computed values
of $E$, s.t. each such $E$-bin contains at least one datum in $\bD$.
\vspace{-.5cm}
\section{Results}
\label{sec:5}
\noindent
We have input data on location and velocities of 2 kinds of galactic particles (called
``Globular Clusters'', and ``Planetary Nebulae'' -- respectively abbreviated as GC and PNe), available for the real galaxy NGC4494. 
The GC data comprises $114$ measurements of $\bY = (X_1,X_2,V_3)^T$,
for the GCs in NGC4494 \cite{GC}. 
Our second data set (PNe data), comprises
$255$ measurements of the PNe \cite{pne}. 
Results of the learnt
$95\%$ HPDs for all parameters, given both PNe (top
row) and GC (bottom row) data, are shown in Figure \ref{fig:HPDs}. Significant inconsistencies between the learnt gravitational mass density parameters can suggest interesting dynamics, such as splitting of the galactic state space into multiple, non-communicating sub-spaces \cite{jasa}, but for this galaxy, it is noted
that such parameters learnt from the 2
datasets, concur within learnt HPDs. 
\section{Conclusions}
\label{sec:6}
\noindent
An astronomical implication of our work is that $\rho_1$ learnt from
either dataset suggests a very high gravitational mass density in the
innermost $R_p$-bin ($\approx$ 1.6kpc), implying gravitational mass $\gtrsim 10^9$times mass of the Sun, enclosed within this innermost radial bin. This
result alone does not contradict the suggestion that NGC4494 harbours a
central supermassive blackhole (SMBH) of mass $\sim 2.69\pm 2.04\times 10^7$ solar masses \cite{blackhole}. Very interestingly, our results indicate that for both GCs and PNe, most particles lie in the intermediate range of energy values; this is also borne by the shape of the histogram of the empirically computed energy using either dataset, where this empirical $E$ value computation is discussed in the last paragraph of Section~\ref{sec:4}. However, owing to its intense radially inward gravitational attraction, a central SMBH is expected to render the potential energy (and therefore the total energy $E$) of the particles closer to the galactic centre, to be much higher negative values, than those further away, while also rendering the number (density) of particles to be sharply monotonically decreasing with radius away from the centre. This is expected to render the energy distribution to be monotonically decreasing as we move towards more positive $E$ values -- in contradiction to our noted non-monotonic trend. So while our results
are not in contradiction to the report of a very large value of mass enclosed within the inner parts of NGC4494, interpretation of that mass as a SMBH does not follow from our learning of the state space $pdf$.

The learning of the gravitational mass density function, and state space $pdf$ -- as well as that of the relation $\bxi(\cdot)$ between the observable state space coordinates, and the system function/vector -- can be done after generating the training dataset relevant to the functional learning problem at hand. Applications in Petrophysics and Finance are also planned.

\renewcommand\baselinestretch{1.}
\small
\bibliographystyle{apalike}


\end{document}